\documentclass[prd,twocolumn,tightenlines,showpacs,nofootinbib]{revtex4-1} 

\usepackage{amsfonts}
\usepackage{dcolumn}
\usepackage{bm}
\usepackage{amsmath}
\usepackage{nicefrac}
\usepackage{amsthm}
\usepackage{amssymb}
\usepackage{graphicx}
\usepackage{color}

\newcommand{\appropto}{\mathrel{\vcenter{
  \offinterlineskip\halign{\hfil$##$\cr
    \propto\cr\noalign{\kern2pt}\sim\cr\noalign{\kern-2pt}}}}}

\begin{document}

\title{Constraining scalar dark matter with Big Bang nucleosynthesis and atomic spectroscopy}

\date{\today}
\author{Y.~V.~Stadnik} \email[]{y.stadnik@unsw.edu.au}
\affiliation{School of Physics, University of New South Wales, Sydney 2052, Australia}
\author{V.~V.~Flambaum} \email[]{v.flambaum@unsw.edu.au}
\affiliation{School of Physics, University of New South Wales, Sydney 2052, Australia}

\begin{abstract}
Scalar dark matter can interact with Standard Model (SM) particles, altering the fundamental constants of Nature in the process. Changes in the fundamental constants during and prior to Big Bang nucleosynthesis (BBN) produce changes in the primordial abundances of the light elements. 
By comparing the measured and calculated (within the SM) primordial abundance of $^{4}$He, which is predominantly determined by the ratio of the neutron-proton mass difference to freeze-out temperature at the time of weak interaction freeze-out prior to BBN, we are able to derive stringent constraints on the mass of a scalar dark matter particle $\phi$ together with its interactions with the photon, light quarks and massive vector bosons via quadratic couplings in $\phi$, as well as its interactions with massive vector bosons via linear couplings in $\phi$. We also derive a stringent constraint on the quadratic interaction of $\phi$ with the photon from recent atomic dysprosium spectroscopy measurements. 

\end{abstract}

\maketitle

\section{Introduction}
\label{Sec:Introduction}
Measurements \cite{Olive2004,Peimbert2007,Izotov2007} and Standard Model (SM) calculations \cite{Wagoner1967,Wetterich2007,Dent2008,Olive2008a,Berengut2010,Epelbaum2013} of the primordial abundances of the light elements produced during Big Bang nucleosynthesis (BBN) provide a valuable probe into physics beyond the SM, including axion-like pseudoscalar dark matter (DM) \cite{Blum2014}, neutrinos and relativistic non-SM particles in the early Universe \cite{Olive2004,Steigman2012}, DM reactions during BBN \cite{Olive2004,Pospelov2009,Pires2014}, variations of the fundamental constants of Nature \cite{Flambaum2002,Shuryak2003,Dmitriev2003,Dmitriev2004,Wetterich2004,Olive2004,Wetterich2007,Dent2008,Coc2007,Landau2009,Berengut2010,Bedaque2011,Epelbaum2013} and supersymmetric models \cite{Steffen2008}. BBN considerations have also been invoked to explain the fine-tuning of the weak scale in the SM \cite{Hall2014}. For contemporary reviews of various aspects of BBN, we refer the reader to Refs.~\cite{Pradler2010,Sarkar2014}.

In our previous work \cite{Stadnik2015DM-VFCs}, we pointed out that a global cosmological evolution of the fundamental constants due to an oscillating scalar DM field $\phi$ that interacts with SM particles via quadratic couplings in $\phi$ is most stringently constrained by determination of the ratio of the neutron-proton mass difference to freeze-out temperature at the time of the weak interaction freeze-out prior to BBN. The stringency of the constraints arises due to the high energy density of DM in the early universe compared with that at present, which enhances the effects of any possible variations of the fundamental constants due to the underlying field $\phi$. 

In the present letter, we derive constraints on the mass of a scalar DM particle $\phi$ together with its interactions with the photon, light quarks and massive vector bosons via quadratic couplings in $\phi$, as well as its interactions with massive vector bosons via linear couplings in $\phi$, by comparing the measured and calculated (within the SM) primordial abundance of $^{4}$He, which is predominantly determined by the ratio of the neutron-proton mass difference to freeze-out temperature at the time of weak interaction freeze-out prior to BBN. We also derive a constraint on the quadratic interaction of $\phi$ with the photon from recent atomic dysprosium spectroscopy measurements. Our constraints are found to rule out large regions of previously unconstrained scalar DM parameter space.

\section{Theory}
\label{Sec:Theory}
In the present work, we consider a scalar DM field $\phi$, which can couple to the SM fields via the following linear-in-$\phi$ interactions:
\begin{equation}
\label{scalar-fermion_osc}
\mathcal{L}_{\textrm{int}}^{f} = \mp \sum_{f}  \frac{\phi}{\Lambda_f} m_f  \bar{f}f ,
\end{equation}
where the sum runs over all SM fermions $f$, $m_f$ is the standard mass of the fermion, $f$ is the fermion Dirac field and $\bar{f}=f^\dagger \gamma^0$, 
\begin{equation}
\label{scalar-photon_osc}
\mathcal{L}_{\textrm{int}}^{\gamma} =  \pm \frac{\phi}{\Lambda_\gamma} \frac{F_{\mu \nu} F^{\mu \nu}}{4} ,
\end{equation}
where $F_{\mu \nu}$ are the components of the electromagnetic field tensor, and
\begin{equation}
\label{scalar-MVB_osc}
\mathcal{L}_{\textrm{int}}^{V} = \pm \sum_{V} \frac{ \phi}{\Lambda_V} \frac{M_V^2}{2} V_\nu V^\nu ,
\end{equation}
where the sum runs over all SM massive vector bosons $V$, $M_V$ is the standard mass of the boson and $V_\nu$ are the components of the wavefunction of the corresponding massive vector boson. $\Lambda_X$ is a very large energy scale, which is strongly constrained by equivalence principle tests, including lunar laser ranging \cite{Turyshev2004,Turyshev2012} and the E\"{o}tWash experiment \cite{Adelberger2008,Adelberger2012} (see also \cite{Raffelt1999} for constraints from stellar energy loss bounds). Eqs.~(\ref{scalar-fermion_osc}), (\ref{scalar-photon_osc}) and (\ref{scalar-MVB_osc}) alter the fundamental constants as follows, respectively:
\begin{equation}
\label{delta-m_f}
m_f \to m_f \left[1 \pm \frac{\phi}{\Lambda_f} \right] ,
\end{equation}
\begin{equation}
\label{delta-alpha}
\alpha \to \frac{\alpha}{1 \mp \phi / \Lambda_\gamma } \simeq \alpha \left[1 \pm \frac{\phi}{\Lambda_\gamma} \right] ,
\end{equation}
\begin{equation}
\label{delta-M_V}
M_V \to M_V \left[1 \pm \frac{\phi}{\Lambda_V} \right] .
\end{equation}
The scalar DM field $\phi$ can also couple to the SM fields via quadratic-in-$\phi$ interactions, with the replacement $\phi/\Lambda_X \to (\phi/\Lambda'_X)^2$ in Eqs.~(\ref{scalar-fermion_osc}) -- (\ref{delta-M_V}). $\Lambda'_X$ is a large energy scale, which is constrained by astrophysical observations, most notably bounds from supernova energy loss, and equivalence principle tests \cite{Olive2008}. 

Scalar DM can thus induce a cosmological evolution of the fundamental constants. Changes in the fundamental constants during and prior to BBN alter the primordial abundances of the light elements. There are two limiting cases of particular interest to consider:

(i) $\phi$ is an oscillating field and interacts via SM particles via quadratic couplings in $\phi$. This occurs when $m_\phi \gg H(t)$, where $m_\phi$ is the mass of the scalar DM particle and $H(t) = 1/2t$ is the Hubble parameter as a function of time \cite{Rubakov_Book}. The energy density of a non-relativistic oscillating scalar DM field is given by $\rho_{\textrm{scalar}} \simeq  m_\phi^2 \left<\phi^2 \right>$ and for a non-relativistic cold field evolves according to the relation
\begin{equation}
\label{rho-temp}
\bar{\rho}_{\textrm{DM}} = 1.3 \times 10^{-6} \left[1 + z(t) \right]^3  \frac{\textrm{GeV}}{\textrm{cm}^3}  ,
\end{equation}
where $z(t)$ is the redshift parameter and the present mean DM energy density is determined from WMAP measurements \cite{PDG2012}. 

(ii) $\phi$ is a non-oscillating field and interacts with SM particles via linear or quadratic couplings in $\phi$. This occurs when $m_\phi \ll H(t)$. The energy density of a non-oscillating scalar DM field is given by $\rho_{\textrm{scalar}} =  m_\phi^2 \left<\phi^2\right> / 2$ and, due to Hubble friction, is approximately constant while the field remains non-oscillating:
\begin{equation}
\label{rho-temp_OD}
\bar{\rho}_{\textrm{DM}} \approx 1.3 \times 10^{-6} \left[1 + z(t_m) \right]^3  \frac{\textrm{GeV}}{\textrm{cm}^3}  ,
\end{equation}
where $z(t_m)$ is defined by $H(t_m) = m_\phi$. 

For constraining the parameters of scalar DM from measurements and SM calculations of the primordial abundance of $^{4}$He, it suffices to consider only the effects of variation of the fundamental constants due to scalar DM on the ratio of the neutron-proton mass difference to freeze-out temperature at the time of weak interaction freeze-out ($t_{\textrm{F}} \approx 1.1$ s), which determines the abundance of neutrons available for BBN (the vast majority of these neutrons are ultimately locked up in $^{4}$He). The corresponding range of scalar DM particle masses for (i) is hence $m_\phi \gg 10^{-16}$ eV, while the corresponding range of scalar DM particle masses for (ii) is $m_\phi \ll 10^{-16}$ eV.

\section{Results}
\label{Sec:Constraints}
The loosest constraints on the parameters of $\phi$ are obtained by assuming that $\phi$ is a non-relativistic cold field at all times, which is produced either non-thermally (through vacuum decay \cite{Preskill1982}) or thermally (with a very large mass). An example of a cold scalar DM field is a classical oscillating condensate, $\phi = \phi_0 \cos(\omega t)$, which oscillates with frequency $\omega \approx m_\phi c^2 / \hbar$. We note that BBN and Cosmic Microwave Background (CMB) measurements do not rule out the existence of a relativistic scalar DM field \cite{Olive2004}, and that a relativistic scalar DM field should provide even more stringent constraints on the underlying parameters than those derived in the present work, since for a relativistic oscillating DM field, the mean DM energy density evolves according to $\bar{\rho}_{\textrm{DM}} \propto [1+z(t)]^4$. 

Constraints on scalar DM parameters follow from measurements and SM calculations of the abundance of $^{4}$He produced during BBN, which depends strongly on the neutron-to-proton ratio at the time of the weak interaction freeze-out ($T_\textrm{F} = b M_W^{4/3} \sin^{4/3}(\theta_\textrm{W}) / (\alpha^{2/3} M_{\textrm{Planck}}^{1/3}) \approx 0.75$ MeV \cite{Rubakov_Book}), where $T_\textrm{F}$ is the weak interaction freeze-out temperature, $\theta_\textrm{W}$ is the Weinberg angle, $\alpha$ is the electromagnetic fine-structure constant, $M_{\textrm{Planck}}$ is the Planck mass and $b$ is a numerical constant. The neutron-to-proton ratio at the time of weak interaction freeze-out is given by
\begin{equation}
\label{np_ratio}
\left(\frac{n}{p}\right)_{\textrm{weak}} = e^{-Q_{np}/ T_\textrm{F}} ,
\end{equation}
where $Q_{np}$ is the neutron-proton mass difference:
\begin{equation}
\label{np_mass_diff}
Q_{np} = m_n - m_p = a\alpha \Lambda_{\textrm{QCD}} + (m_d - m_u)  , 
\end{equation}
with the present-day values $(a\alpha \Lambda_{\textrm{QCD}})_0 = -0.76$ MeV, where $\Lambda_{\textrm{QCD}}$ is the Quantum Chromodynamics scale and $a$ is a numerical constant, and $(m_d-m_u)_0 = 2.05$ MeV \cite{Leutwyler1982}. From the measured and predicted (within the SM) primordial $^{4}$He abundance, $Y_p^{\textrm{exp}} (^4 \textrm{He}) = 0.2474 \pm 0.0028$ \cite{Peimbert2007} and $Y_p^{\textrm{theor}} (^4 \textrm{He}) = 0.2486 \pm 0.0002$ \cite{Berengut2010}, we find the measured and predicted $n/p$ ratio at the time of BBN freeze-out to be
\begin{equation}
\label{n/p_exp_BBN}
\left(\frac{n}{p}\right)_{\textrm{BBN}}^{\textrm{exp}} = 0.1420 \pm 0.0021 ,
\end{equation}
\begin{equation}
\label{n/p_theor_BBN}
\left(\frac{n}{p}\right)_{\textrm{BBN}}^{\textrm{theor}} = 0.1428 \pm 0.0001 .
\end{equation}
Extrapolating back to the time of weak interaction freeze-out ($\Delta t \approx 180$ s) with a neutron half-life of $\tau_n = 880$ s \cite{PDG2012}, gives the measured and predicted $n/p$ ratio at the time of weak interaction freeze-out:
\begin{equation}
\label{n/p_exp_weak}
\left(\frac{n}{p}\right)_{\textrm{weak}}^{\textrm{exp}} = 0.1801 \pm 0.0026 ,
\end{equation}
\begin{equation}
\label{n/p_theor_weak}
\left(\frac{n}{p}\right)_{\textrm{weak}}^{\textrm{theor}} = 0.1811 \pm 0.0002 ,
\end{equation}
which correspond to the following measured and predicted $Q_{np}/T_{\textrm{F}}$ ratio:
\begin{equation}
\label{Q_np_exp_weak}
\left(\frac{Q_{np}}{T_\textrm{F}}\right)^{\textrm{exp}} = 1.714 \pm 0.015 ,
\end{equation}
\begin{equation}
\label{Q_np_theor_weak}
\left(\frac{Q_{np}}{T_\textrm{F}}\right)^{\textrm{theor}} = 1.709 \pm 0.001 .
\end{equation}
The relative difference in the measured and predicted $Q_{np}/T_{\textrm{F}}$ ratio is found to be (adding the experimental and theoretical uncertainties in quadrature):
\begin{equation}
\label{Q_np_rel_diff}
\frac{\Delta (Q_{np}/T_\textrm{F})}{Q_{np}/T_\textrm{F}} = 0.0033 \pm 0.0085 .
\end{equation}
(\ref{Q_np_rel_diff}) can be interpretted as a constraint on temporal variations in the underlying fundamental constants from the time of weak interaction freeze-out until the present time:
\begin{align}
\label{constraints_WIFO_1}
 0.08 \frac{\Delta \alpha}{\alpha} + 1.59 \frac{\Delta (m_d - m_u)}{(m_d - m_u)} +3.32 \frac{\Delta M_W}{M_W} - 4.65 \frac{\Delta M_Z}{M_Z}  \notag \\
-0.59 \frac{\Delta \Lambda_{\textrm{QCD}}}{\Lambda_{\textrm{QCD}}} + \frac{1}{3} \frac{\Delta M_{\textrm{Planck}}}{M_{\textrm{Planck}}} = 0.0033 \pm 0.0085 ,
\end{align}
where we have made use of the relation $\cos(\theta_\textrm{W}) = M_W/M_Z = 0.882$ \cite{PDG2012}. 

\begin{figure}[h!]
\begin{center}
\includegraphics[width=8.5cm]{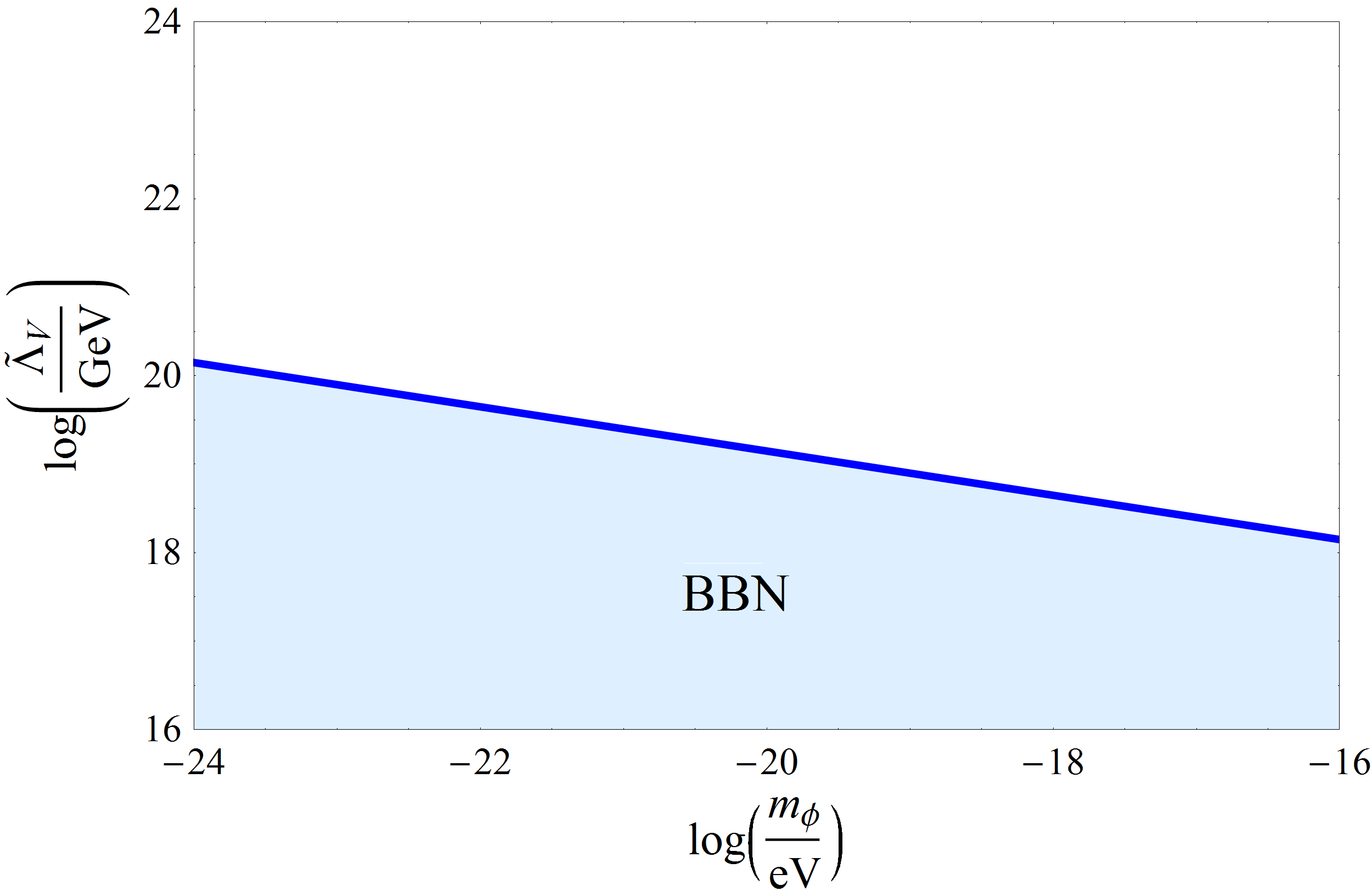}
\caption{(Color online) Region of scalar dark matter parameter space ruled out for the linear interaction of $\phi$ with massive vector bosons. Region below blue line corresponds to constraints derived in the present work from comparison of measurements and SM calculations of the ratio $Q_{np}/ T_\textrm{F}$, for the case $\tilde{\Lambda}_{V} \ll \Lambda_\gamma$, $\tilde{\Lambda}_{V} \ll |\Lambda_u \Lambda_d (m_d - m_u) / (\Lambda_u m_d - \Lambda_d m_u)|$.} 
\label{fig:Lambda_vector_linear_space}
\end{center}
\end{figure}

The constraint (\ref{constraints_WIFO_1}) can be expressed in terms of $\phi$ and $\Lambda_X$ as follows (retaining only variations in the fundamental constants that are induced by the linear-in-$\phi$ interactions considered in the present work):
\begin{align}
\label{constraints_WIFO_2a+L}
\left[\left<\phi\right>_{\textrm{weak}} - \left<\phi\right>_0 \right] \left[ \frac{0.08 \kappa_\gamma}{\Lambda_\gamma} + \frac{1.59}{m_d-m_u} \left(\frac{\kappa_d m_d}{\Lambda_d} - \frac{\kappa_u m_u}{\Lambda_u} \right)   \right. \notag \\ 
\left. + \frac{3.32 \kappa_W}{\Lambda_W} - \frac{4.65 \kappa_Z}{\Lambda_Z} \right] =  0.0033 \pm 0.0085 ,
\end{align}
where $\kappa_X = \pm 1$ correspond to the relevant signs in the Lagrangians (\ref{scalar-fermion_osc}) -- (\ref{scalar-MVB_osc}). 
For a non-oscillating scalar DM field ($m_\phi \ll 10^{-16}$ eV at $t_{\textrm{F}} \approx 1.1$ s), for which the energy density evolves according to (\ref{rho-temp_OD}), this leads to
\begin{widetext}
\begin{align}
\label{constraints_WIFO_3a++L}
\frac{1}{m_\phi} \left(\frac{m_\phi}{3 \times 10^{-16} ~\textrm{eV}}\right)^{3/4}  \left[ \frac{0.08 \kappa_\gamma}{\Lambda_\gamma} + \frac{1.59}{m_d-m_u} \left(\frac{\kappa_d m_d}{\Lambda_d} - \frac{\kappa_u m_u}{\Lambda_u} \right)     + \frac{3.32 \kappa_W}{\Lambda_W}  - \frac{4.65 \kappa_Z}{\Lambda_Z} \right] \simeq  (0.4 \pm 1.0) \times 10^{-11} ~\textrm{eV}^{-2} ,
\end{align}
\end{widetext}
where we have made use of the fact that $|\left<\phi\right>_{\textrm{weak}}| \gg |\left<\phi\right>_0|$ and the relation $[1+z(t_m)] / (1+z_{\textrm{F}}) \simeq \sqrt{t_{\textrm{F}}/t_m}$ during and after BBN (but at much earlier times than electron-proton recombination), and assumed that scalar DM saturates the present-day DM energy density. We note that a single type of measurement does not give constraints on the individual parameters appearing in (\ref{constraints_WIFO_3a++L}), but rather gives constraints on a combination thereof. However, we can extract useful information about the underlying parameters by sequentially assuming that individual terms within (\ref{constraints_WIFO_3a++L}) dominate the others. The resulting region of parameter space excluded by comparison of measurements and SM calculations of the primordial abundance of $^{4}$He is shown in Fig.~\ref{fig:Lambda_vector_linear_space}, in terms of the parameter $\tilde{\Lambda}_{V} = |\Lambda_W \Lambda_Z / (\Lambda_Z - 1.40 \Lambda_W)|$, assuming $\kappa_W= \kappa_Z$. For the linear interaction of $\phi$ with the photon and light quarks, the BBN limits are weaker than existing limits from equivalence principle tests \cite{Adelberger2008,Adelberger2012} and dysprosium spectroscopy measurements \cite{Budker2015} in the mass range of interest.

\begin{figure}[h!]
\begin{center}
\includegraphics[width=8.5cm]{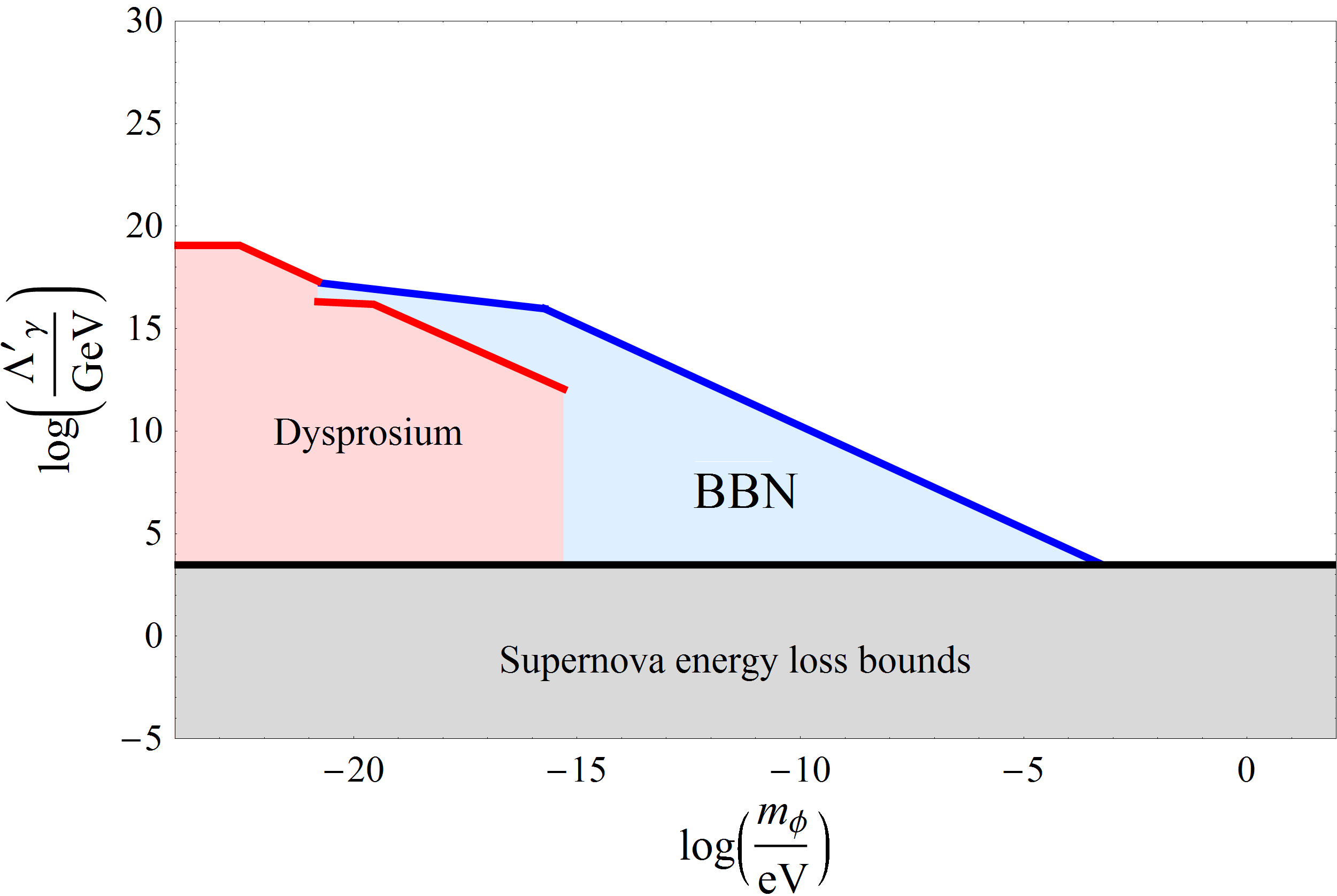}
\caption{(Color online) Region of scalar dark matter parameter space ruled out for the quadratic interaction of $\phi$ with the photon. Region below blue line corresponds to constraints derived in the present work from comparison of measurements and SM calculations of the ratio $Q_{np}/ T_\textrm{F}$, for the case $\Lambda'_\gamma \ll \tilde{\Lambda}'_{q}$, $\Lambda'_\gamma \ll \tilde{\Lambda}'_{V}$. Region below black line corresponds to constraints from supernova energy loss bounds \cite{Olive2008}. Region below red line corresponds to constraints derived in the present work from the data of recent atomic dysprosium spectroscopy measurements in \cite{Budker2015}.} 
\label{fig:Lambda_gamma_quadratic_space}
\end{center}
\end{figure}

\begin{figure}[h!]
\begin{center}
\includegraphics[width=8.5cm]{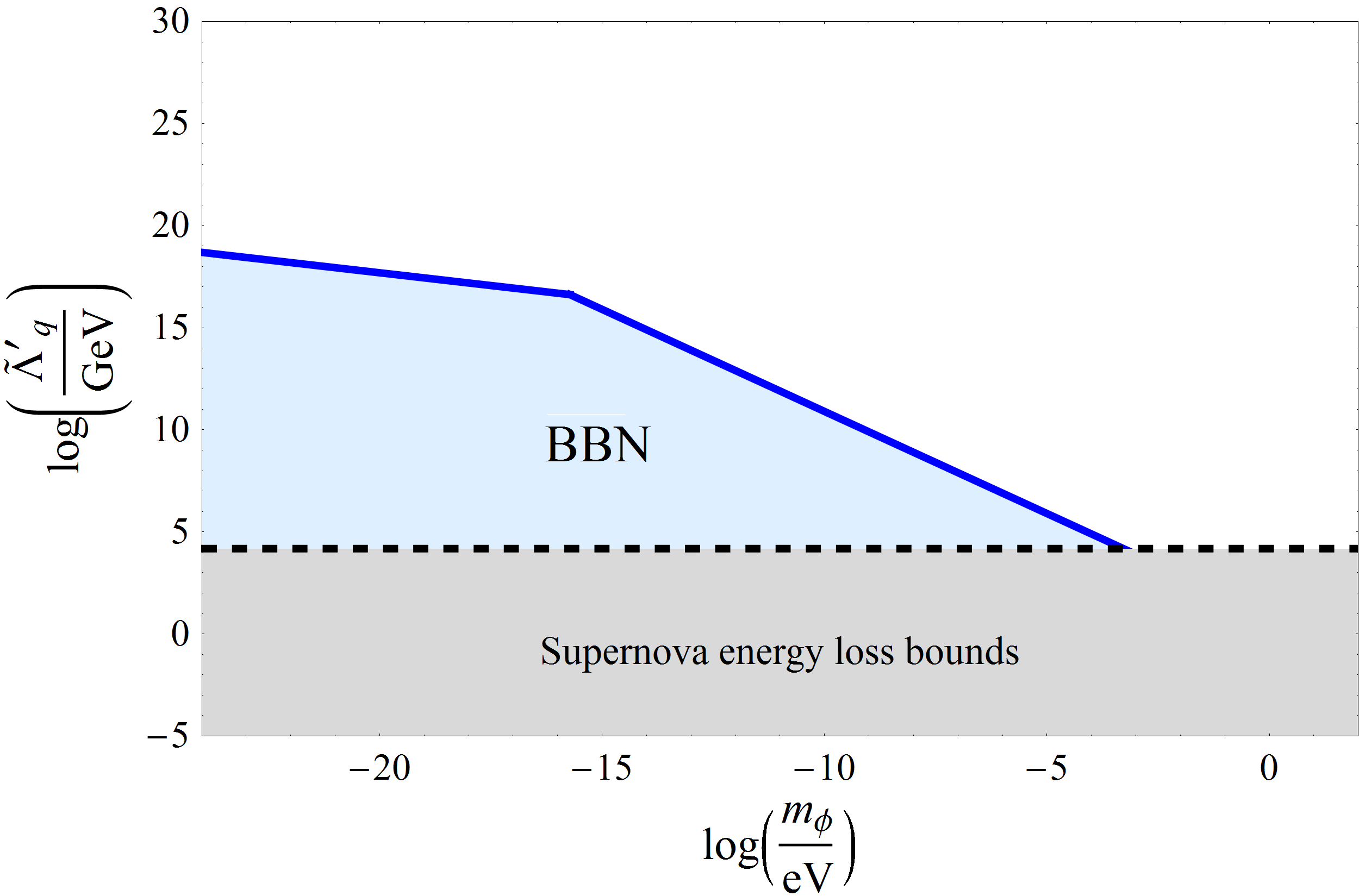}
\caption{(Color online) Region of scalar dark matter parameter space ruled out for the quadratic interaction of $\phi$ with light quarks. Region below blue line corresponds to constraints derived in the present work from comparison of measurements and SM calculations of the ratio $Q_{np}/ T_\textrm{F}$, for the case $\tilde{\Lambda}'_{q} \ll \Lambda'_\gamma$, $\tilde{\Lambda}'_{q} \ll \tilde{\Lambda}'_{V}$. Region below dashed black line corresponds to constraints on $\Lambda'_p$ from supernova energy loss bounds \cite{Olive2008}. Strictly, the limits are independent, since they contain different linear combinations of the quark interaction parameters.} 
\label{fig:Lambda_quark_quadratic_space}
\end{center}
\end{figure}

\begin{figure}[h!]
\begin{center}
\includegraphics[width=8.5cm]{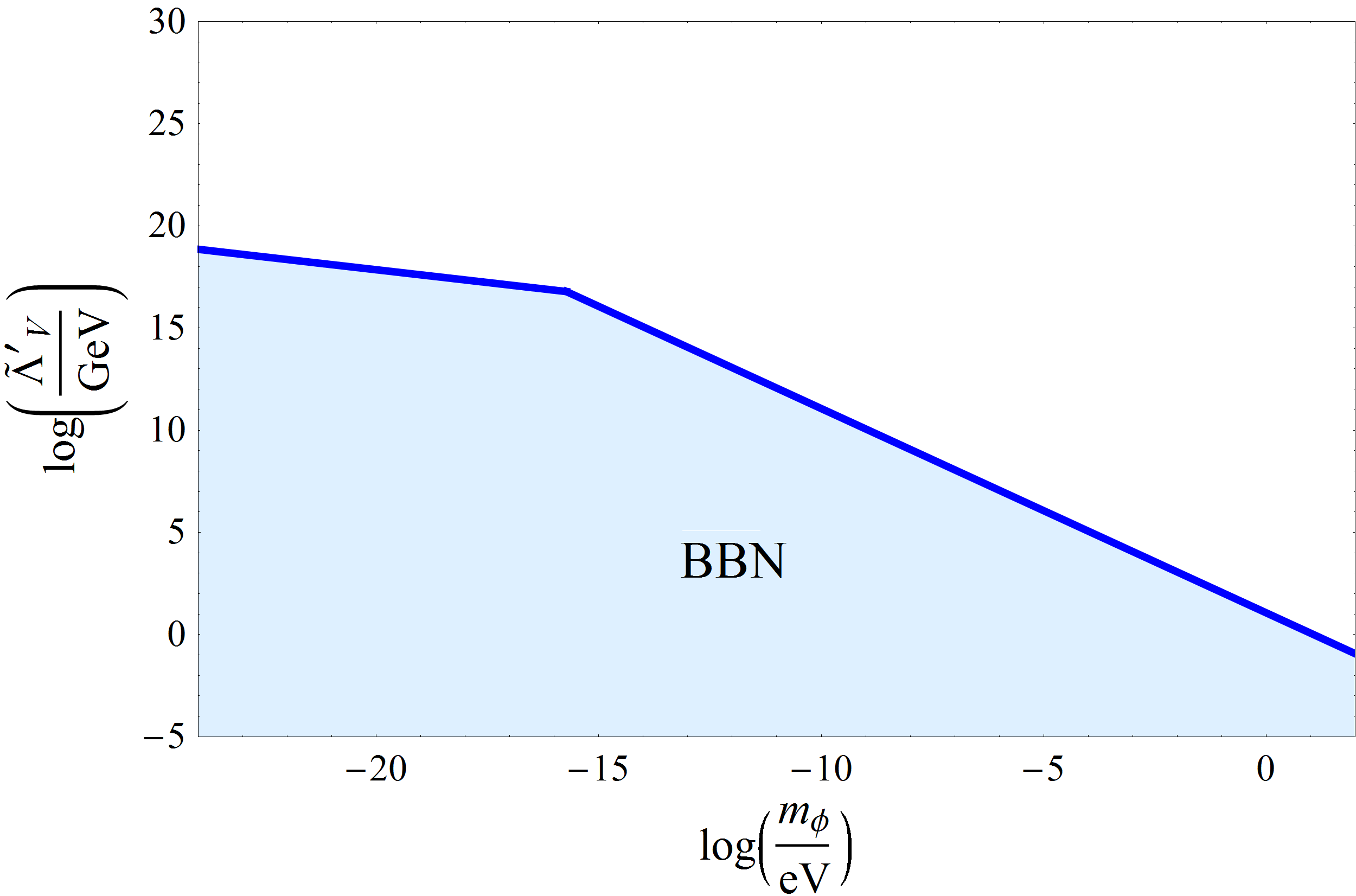}
\caption{(Color online) Region of scalar dark matter parameter space ruled out for the quadratic interaction of $\phi$ with massive vector bosons. Region below blue line corresponds to constraints derived in the present work from comparison of measurements and SM calculations of the ratio $Q_{np}/ T_\textrm{F}$, for the case $\tilde{\Lambda}'_{V} \ll \Lambda'_\gamma$, $\tilde{\Lambda}'_{V} \ll \tilde{\Lambda}'_{q}$.} 
\label{fig:Lambda_vector_quadratic_space}
\end{center}
\end{figure}

The constraint (\ref{constraints_WIFO_1}) can also be expressed in terms of $\phi$ and $\Lambda'_X$ as follows (retaining only variations in the fundamental constants that are induced by the quadratic-in-$\phi$ interactions considered in the present work):
\begin{align}
\label{constraints_WIFO_2a}
\left[\left<\phi^2\right>_{\textrm{weak}} - \left<\phi^2\right>_0 \right] \left[ \frac{0.08 \kappa'_\gamma}{(\Lambda_\gamma')^2} + \frac{1.59}{m_d-m_u} \left(\frac{\kappa'_d m_d}{(\Lambda'_d)^2} - \frac{\kappa'_u m_u}{(\Lambda'_u)^2} \right)   \right. \notag \\ 
\left. + \frac{3.32 \kappa'_W}{(\Lambda'_W)^2} - \frac{4.65 \kappa'_Z}{(\Lambda'_Z)^2} \right] =  0.0033 \pm 0.0085 .
\end{align}
For an oscillating scalar DM field ($m_\phi \gg 10^{-16}$ eV at $t_{\textrm{F}} \approx 1.1$ s), for which the energy density evolves according to (\ref{rho-temp}), this leads to
\begin{align}
\label{constraints_WIFO_3a}
\frac{1}{m_\phi^2}   \left[ \frac{0.08 \kappa'_\gamma}{(\Lambda'_\gamma)^2} + \frac{1.59}{m_d-m_u} \left(\frac{\kappa'_d m_d}{(\Lambda'_d)^2} - \frac{\kappa'_u m_u}{(\Lambda'_u)^2} \right) + \frac{3.32 \kappa'_W}{(\Lambda'_W)^2}  \right. \notag \\ 
\left.  - \frac{4.65 \kappa'_Z}{(\Lambda'_Z)^2} \right] \simeq  (1.0 \pm 2.5) \times 10^{-20} ~\textrm{eV}^{-4} ,
\end{align}
where we have made use of the fact that $\left<\phi^2\right>_{\textrm{weak}} \gg \left<\phi^2\right>_0$ and assumed that scalar DM saturates the present-day DM energy density. 
For a non-oscillating scalar DM field ($m_\phi \ll 10^{-16}$ eV at $t_{\textrm{F}} \approx 1.1$ s), for which the energy density evolves according to (\ref{rho-temp_OD}), this leads to
\begin{widetext}
\begin{align}
\label{constraints_WIFO_3a++}
\frac{1}{m_\phi^2} \left(\frac{m_\phi}{3 \times 10^{-16} ~\textrm{eV}}\right)^{3/2}  \left[ \frac{0.08 \kappa'_\gamma}{(\Lambda'_\gamma)^2} + \frac{1.59}{m_d-m_u} \left(\frac{\kappa'_d m_d}{(\Lambda'_d)^2} - \frac{\kappa'_u m_u}{(\Lambda'_u)^2} \right)    + \frac{3.32 \kappa'_W}{(\Lambda'_W)^2}  - \frac{4.65 \kappa'_Z}{(\Lambda'_Z)^2} \right] \simeq  (0.5 \pm 1.3) \times 10^{-20} ~\textrm{eV}^{-4} ,
\end{align}
\end{widetext}
where we have made use of the fact that $\left<\phi^2\right>_{\textrm{weak}} \gg \left<\phi^2\right>_0$ and the relation $[1+z(t_m)] / (1+z_{\textrm{F}}) \simeq \sqrt{t_{\textrm{F}}/t_m}$ during and after BBN (but at much earlier times than electron-proton recombination), and assumed that scalar DM saturates the present-day DM energy density. The resulting regions of parameter space excluded by comparison of measurements and SM calculations of the primordial abundance of $^{4}$He are shown in Figs.~\ref{fig:Lambda_gamma_quadratic_space} -- \ref{fig:Lambda_vector_quadratic_space}, in terms of the parameters $\Lambda'_\gamma$, $(\tilde{\Lambda}'_{q})^2 = |(\Lambda'_u)^2 (\Lambda'_d)^2 (m_d - m_u) / [(\Lambda'_u)^2 m_d - (\Lambda'_d)^2 m_u]|$ and $(\tilde{\Lambda}'_{V})^2 = |(\Lambda'_W)^2 (\Lambda'_Z)^2 / [(\Lambda'_Z)^2 - 1.40 (\Lambda'_W)^2]|$, assuming $\kappa'_d= \kappa'_u$ and $\kappa'_W= \kappa'_Z$. Note that we have extrapolated the constraints from the two limiting mass ranges to the connecting central region. 

We have also derived constraints on the quadratic interaction of $\phi$ with the photon, using data from recent atomic dysprosium spectroscopy measurements in Ref.~\cite{Budker2015} pertaining to oscillating variations in $\alpha$. These constraints are presented in Fig.~\ref{fig:Lambda_gamma_quadratic_space}.

\section{Conclusions}
\label{Sec:Conclusions}
By comparing the measured and calculated (within the Standard Model) primordial abundance of $^{4}$He, which is predominantly determined by the ratio of the neutron-proton mass difference to freeze-out temperature at the time of weak interaction freeze-out prior to Big Bang nucleosynthesis, we have derived stringent constraints on the mass of a scalar dark matter particle $\phi$ together with its interactions with the photon, light quarks and massive vector bosons via quadratic couplings in $\phi$, as well as its interactions with massive vector bosons via linear couplings in $\phi$. We have also derived constraints on the quadratic interaction of $\phi$ with the photon from recent atomic dysprosium spectroscopy measurements. The new constraints derived in the present work rule out large regions of previously unconstrained scalar dark matter parameter space.

\section*{ACKNOWLEDGEMENTS}
We would like to thank Ken Van Tilburg for helpful discussions. This work was supported by the Australian Research Council.




\end{document}